\documentclass[pra,showpacs,twocolumn]{revtex4}

\usepackage{amsmath,amssymb,amsfonts,graphicx}

\renewcommand{\vec}[1]{\mathbf{#1}}
\newcommand{\ket}[1]{\left|#1\right>}
\newcommand{\bra}[1]{\left<#1\right|}
\newcommand{\rep}[1]{\text{\textsf{#1}}}

\begin{document}

\title{Lorentz--covariant reduced spin density matrix and Einstein--Podolsky--Rosen--Bohm
  correlations} 

\author{Pawe{\l} Caban}\email{P.Caban@merlin.fic.uni.lodz.pl}
\author{Jakub
  Rembieli\'nski}\email{J.Rembielinski@merlin.fic.uni.lodz.pl}
\affiliation{Department of Theoretical Physics,
University of Lodz\\ 
Pomorska 149/153, 90-236 {\L}\'od\'z, Poland}

\date{\today}

\begin{abstract}
We show that it is possible to define a Lorentz--covariant reduced spin
density matrix for massive particles. Such a matrix allows one to
calculate the mean 
values of observables connected with spin measurements (average
polarizations). 
Moreover, it contains not only information about polarization of the
particle but also information about its \textit{average}
kinematical state. We also use our formalism to calculate
the correlation function in the Einstein--Podolsky--Rosen--Bohm type
experiment with massive relativistic particles. 
\end{abstract}


\maketitle

\section{Introduction}

Relativistic aspects of quantum mechanics have recently
attracted much attention, especially in the context of the theory of
quantum information. One of the important questions in this
context is how to define the reduced spin density matrix. 
Such a matrix should enable one to make statistical predictions for the
outcomes of ideal spin measurements which are not influenced by the
particle momentum. 
We consider
this problem in detail in the case of massive 
particles.  The reduced spin density matrix is usually defined by the
following formula \cite{cab_PST2002}:
 \begin{equation}
 \tau_{\sigma\lambda}=\int d\mu(k)
 \bra{k,m,s,\sigma}\hat{\rho}\ket{k,m,s\lambda}, 
 \label{red_density_standard}
 \end{equation}
where $\hat{\rho}$ denotes the complete density matrix of a
single particle with mass $m$,
$d\mu(k)=\frac{d^3k}{2k^0}$ is 
the Lorentz--invariant measure on the mass shell 
and four-momentum eigenvectors $\ket{k,m,s,\lambda}$
(i.e., $P_\mu\ket{k,m,s,\lambda}=k_\mu\ket{k,m,s,\lambda}$) span the
space of the 
irreducible representation of the Poincar\'e group. They are
normalized as follows
 \begin{equation}
 \left<p,m,s,\sigma|k,m,s,\lambda\right>=2k^0\delta^3(\vec{k}-\vec{p})
 \delta_{\sigma\lambda}.   
 \label{normalization_standard}
 \end{equation}  
The action of the Lorentz transformation $\Lambda$ on the vector
$\ket{k,m,s,\lambda}$ is of the form
 \begin{equation}
 U(\Lambda)\ket{k,m,s,\lambda}=
 {\mathcal{D}}_{\sigma\lambda}^{s}(R(\Lambda,k)) \ket{\Lambda
 k,m,s,\sigma},  
 \label{transf_states_standard}
 \end{equation}
where ${\mathcal{D}}^s$ is the matrix spin $s$ representation of the $SO(3)$
group, $R(\Lambda,k)=L_{\Lambda k}^{-1}\Lambda L_k$ is the Wigner
rotation and $L_k$ designates the standard Lorentz boost defined by
the relations $L_k\tilde{k}=k$, $L_{\tilde{k}}=I$,
$\tilde{k}=(m,\vec{0})$.  

The key question is whether the reduced density matrix is
covariant. In \cite{cab_PST2002} it was stressed that the matrix
(\ref{red_density_standard}) is not covariant under Lorentz boosts. It
means that when we calculate the complete density matrix as seen by
the boosted observer
 \begin{equation}
 \hat{\rho}^\prime=U(\Lambda)\hat{\rho}U^\dag(\Lambda)
 \label{complete_matrix_transf}
 \end{equation} 
and then the reduced spin density matrix
$\tau_{\sigma\lambda}^{\prime}$ (using Eq.\ (\ref{red_density_standard})
with $\hat{\rho}$ replaced with $\hat{\rho}^\prime$) we find that we
cannot express $\tau^\prime$ only in terms of $\tau$ and $\Lambda$. The
reason is quite obvious --- the Wigner rotation in the transformation
law (\ref{transf_states_standard}) is momentum dependent, except of
the case $\Lambda\in O(3)$. 
From the group theoretical point of view it is related to the fact that
the Lorentz group and the
rotation group are not homomorphic. Notice that in the nonrelativistic
quantum mechanics it 
is possible to define the Galilean--covariant reduced density matrix
by the formula analogous to Eq.\ (\ref{red_density_standard})
\cite{cab_CSW2003} because such a homomorphism exists.

\section{Covariant reduced density matrix}

As was pointed out in \cite{cab_Czachor2005} matrix
(\ref{red_density_standard}) is not always relevant to the discussion
of relativistic aspects of polarization experiments (see, however,
\cite{cab_PST2005}). For this reason 
we propose here another definition of the reduced density matrix. This
definition relies on the analogy with the polarization tensors
formalism used in quantum field theory. 
As a result we obtain the finite--dimensional matrix which contains
not only the information about the polarization of the particle but also  
the information about \textit{average} values of the kinematical degrees
of freedom. Moreover, such a matrix transforms
covariantly under the Lorentz group action. 

To begin with we introduce vectors
$\ket{\alpha,k}$ such that
 \begin{equation}
 \ket{\alpha,k}=v_{\alpha\sigma}(k)\ket{k,m,s,\sigma}
 \label{states_new}
 \end{equation}
which are assumed to transform under Lorentz transformation $\Lambda$
due to the following, manifestly covariant, rule
 \begin{equation}
 U(\Lambda)\ket{\alpha,k}=\rep{D}(\Lambda^{-1})_{\alpha\beta}
 \ket{\beta,\Lambda k},
 \label{transf_states_new}
 \end{equation}
where $\rep{D}(\Lambda)$ is a given finite--dimensional Lorentz group
representation. Consistency of the rules
(\ref{transf_states_standard}, \ref{states_new},
\ref{transf_states_new}) leads to the 
Weinberg--like condition \cite{cab_Weinberg1964_1,cab_Weinberg1964_2}
which has to be fulfilled: 
 \begin{equation}
 \rep{D}(\Lambda)v(k){{\mathcal{D}}^s}^T(R(\Lambda,k))=v(\Lambda k),
 \label{Weinberg_condition}
 \end{equation}
where $v(k)$ denotes matrix $[v_{\alpha\sigma}(k)]$. Thus to calculate
$v(k)$ it is enough to determine $v(\tilde{k})$ and use the formula
$v(k)=\rep{D}(L_k)\tilde{k}$ which is a consequence of Eq.\
(\ref{Weinberg_condition}).

Assuming that the condition (\ref{Weinberg_condition}) can be solved
we can define the following (unnormalized) covariant reduced density
matrix:
 \begin{equation}
 \theta_{\alpha\beta}=\int d\mu(k)
 \bra{\beta,k}\hat{\rho}\ket{\alpha,k}.  
 \label{reduced_density_new}
 \end{equation}
We can easily check that this matrix is manifestly
covariant under 
the transformation (\ref{complete_matrix_transf}), namely we have
 \begin{equation}
 \theta^\prime= \rep{D}(\Lambda) \theta \rep{D}^\dag(\Lambda),
 \label{reduced_density_new_transf}
 \end{equation}
where $\theta=[\theta_{\alpha\beta}]$.

One can also easily verify that the matrix
(\ref{reduced_density_new}) is Hermitian and positive
semidefinite (similarly as (\ref{red_density_standard})).
Transformation
(\ref{reduced_density_new_transf}) preserves Hermicity and positive
semidefiniteness of $\theta$ but changes its trace. 

It is clear that we can define also normalized density matrix
 \begin{equation}
 \tilde{\theta}=\frac{\theta}{\text{Tr}\,\theta}.
 \end{equation}
Such a matrix transforms according to the rule
 \begin{equation}
 \tilde{\theta}^\prime =
 \frac{\rep{D}(\Lambda)\tilde{\theta}\rep{D}^\dag(\Lambda}%
 {\text{Tr}\,(\tilde{\theta}\rep{D}^\dag(\Lambda)\rep{D}(\Lambda))}. 
 \label{transf_normalized}
 \end{equation}
One can check immediately that Eq.\ (\ref{transf_normalized}) gives a
nonlinear realization of the Lorentz group connected with the quotient
space $SO(1,3)_0/SO(3)$. Therefore this realization is linear
on the rotation group. However, to extract information about
polarization of the particle it does not matter which matrix we use,
$\theta$ or $\tilde{\theta}$. 
Moreover, when we consider representations of the full Lorentz
group (i.e., including inversions)
the most convenient choice is to consider the matrix 
 \begin{equation}
 \Omega=\theta \Gamma,
 \label{Omega_def}
 \end{equation}
where $\Gamma$ fulfills the condition
 \begin{equation}
 \rep{D}^\dag\Gamma=\Gamma\rep{D}^{-1},
 \label{Gamma_def}
 \end{equation}
which means that in this representation $\Gamma$ represents space
inversions.
Thus the matrix $\Omega$ transforms
under the Lorentz group action in the following way:
 \begin{equation}
 \Omega^\prime = \rep{D}(\Lambda)\Omega\rep{D}^{-1}(\Lambda).
 \label{transf_Omega}
 \end{equation}
We see that transformation (\ref{transf_Omega}) does not change the
trace of $\Omega$. Of course, having $\Omega$ we can easily
determine $\theta$ and normalized density matrix $\tilde{\theta}$.

Hereafter we restrict ourselves to the case of a spin-1/2 particle;
generalization to the higher spin is immediate.
In this case the Weinberg condition
(\ref{Weinberg_condition}) can be easily solved. 
We want to consider representations of the full Lorentz group
thus we choose as the representation $\rep{D}$
the bispinor representation
$D^{(\frac{1}{2},0)}\oplus D^{(0,\frac{1}{2})}$, so $\Gamma=\gamma^0$
in this case. Explicitly, if $A\in
SL(2,{\mathbb{C}})$ and $\Lambda(A)$ is an image of $A$ in the
canonical homomorphism
of the $SL(2,{\mathbb{C}})$ group onto the
Lorentz group, we take the chiral form of $D^{(\frac{1}{2},0)}\oplus
D^{(0,\frac{1}{2})}$, namely 
 \begin{equation}
 \rep{D}(\Lambda(A))= \begin{pmatrix}
 A & 0 \\ 0 & (A^\dag)^{-1}
 \end{pmatrix}.
 \label{representation}
 \end{equation}
The canonical homomorphism between the group
$SL(2,{\mathbb{C}})$ (universal covering of the proper ortochronous
Lorentz group $L_{+}^{\uparrow}$) and the Lorentz group
$L_{+}^{\uparrow}\sim SO(1,3)_0$ \cite{cab_BR1977}  
is defined as follows:
With every four-vector $k^\mu$ we associate a two-dimensional
hermitian matrix ${\textsf k}$ such that
 \begin{equation}
 {\textsf k}=k^\mu \sigma_\mu,
 \label{matrix_k}
 \end{equation}
where $\sigma_i$, $i=1,2,3$, are the standard Pauli matrices
and $\sigma_0=I$. In the space of two-dimensional hermitian matrices
(\ref{matrix_k}) the Lorentz group action is given by 
${\sf k}^\prime=A{\textsf k}A^\dag$, where $A$ denotes the element of the 
$SL(2,{\mathbb{C}})$ group corresponding to the Lorentz transformation 
$\Lambda(A)$ which converts the four-vector $k$ to $k^\prime$ (i.e.,
${k^\prime}^\mu=\Lambda_{\phantom{\mu}\nu}^{\mu}k^\nu$) and
${\textsf k}^\prime={k^\prime}^\mu\sigma_\mu$. 

Now, the explicit solution of the Weinberg condition
(\ref{Weinberg_condition}) under our choice of $\rep{D}$ (Eq.\
(\ref{representation})) is given by
 \begin{equation}
 v(k)=\frac{1}{2\sqrt{1+\frac{k^0}{m}}}\begin{pmatrix}
 (I+\tfrac{1}{m}{\textsf k})\sigma_2 \\
 (I+\tfrac{1}{m}{\textsf k}^P)\sigma_2
 \end{pmatrix},
 \label{amplitude}
 \end{equation}
where ${\textsf k}$ is given by Eq.\ (\ref{matrix_k}) and ${\textsf
  k}^P=(k^P)^\mu \sigma_\mu$ with $k^P=(k^0,-\vec{k})$. 
As is well known, the intertwining
matrix $v(k)$ fulfills the Dirac equation 
 \begin{equation}
 (k\gamma-mI)v(k)=0,
 \end{equation}
where $\gamma^\mu$ are Dirac matrices, $k\gamma=k_\mu
\gamma^\mu$. The explicit representation of Dirac matrices used in the
present paper is summarized in Appendix A.

Now we discuss the general structure of the reduced density
matrix (\ref{reduced_density_new}) for $s=\tfrac{1}{2}$. 
We show that this matrix contains information about
both average 
polarization as well as kinematical degrees of freedom. Recall that
the polarization
of the relativistic particle is determined by the Pauli--Lubanski
four-vector
 \begin{equation}
 W^\mu=\tfrac{1}{2}\varepsilon^{\mu\nu\sigma\lambda}P_\nu
 J_{\sigma\lambda},
 \label{Pauli-Lubanski}
 \end{equation}
where $P_\nu$ is a four-momentum operator and $J_{\sigma\lambda}$
denotes generators of the Lorentz group, i.e.,
$U(\Lambda)=\exp{i\omega^{\mu\nu}J_{\mu\nu}}$. 
We will also use the spin tensor $S_{\mu\nu}$ defined by the
formula \cite{cab_Anderson1967} 
 \begin{equation}
 S_{\mu\nu}= -\frac{1}{m^2}\varepsilon_{\mu\nu\sigma\tau}P^\sigma
 W^\tau. 
 \label{spin_tensor}
 \end{equation}
Now, the $4\times4$ reduced spin density matrix $\theta$ can be
written as the following combination 
 \begin{multline}
 \theta= \frac{1}{4}\big(a \gamma^0+b i\gamma^5\gamma^0+
 u_\mu \gamma^\mu\gamma^0 + \tfrac{2w_\mu}{m}
 \gamma^5\gamma^\mu\gamma^0 \\
 +2 s_{\mu\nu}\tfrac{i}{4}[\gamma^\mu,\gamma^\nu]\gamma^0\big).
 \end{multline}
Real coefficients $a$, $b$, $u_\mu$, $w_\mu$, $s_{\mu\nu}$ can be determined  
by calculating corresponding traces. Thus, after some algebra, using Eqs.\
(\ref{reduced_density_new}, \ref{Omega_def},
\ref{amplitude}--\ref{spin_tensor}) and 
(\ref{normalization}--\ref{additional_relations}) we get
 \begin{align}
 a &= {\text{Tr}}\,(\Omega)=1,\\
 b &={\text{Tr}}\,(i\Omega\gamma_5)=0\\
 u_\mu &=\text{Tr}\,(\Omega\gamma_\mu)
 =\tfrac{1}{m}\left<P_\mu\right>_{\hat{\rho}},\label{average_P} \\ 
 w_\mu &=\frac{m}{2}\text{Tr}\,(\Omega\gamma_\mu\gamma_5) =
 \left<W_\mu\right>_{\hat{\rho}},\\ 
 s_{\mu\nu} &=\text{Tr}\,(\Omega\tfrac{i}{4}[\gamma_\mu,\gamma_\nu])=
 \left<S_{\mu\nu}\right>_{\hat{\rho}}, 
 \end{align}
where $\left<A\right>_{\hat{\rho}}$ denotes the mean value of the
observable $A$ in the state described by the complete density matrix
$\hat{\rho}$, $\left<A\right>_{\hat{\rho}}=\text{Tr}\,(\hat{\rho}A)$. 
Notice that the above relations are not accidental, since
$\gamma^0\gamma^\mu$ is a canonical four-velocity operator for the Dirac
particle and $\tfrac{i}{4}[\gamma^\mu,\gamma^\nu]$ are Lorentz group
generators in the bispinor representation. Thus, finally, the matrix
$\Omega=\theta\gamma^0$ has the following form:
 \begin{multline}
 \Omega= \tfrac{1}{4}I + \tfrac{1}{4m}\left<P_\mu\right>_{\hat{\rho}}
 \gamma^\mu + 
 \tfrac{1}{2m}\left<W_\mu\right>_{\hat{\rho}} \gamma^5 \gamma^\mu \\
 +\tfrac{1}{2} \left<S_{\mu\nu}\right>_{\hat{\rho}}
 \tfrac{i}{4}[\gamma^\mu,\gamma^\nu]. 
 \label{Omega_final}
 \end{multline}

It can be also checked that in the nonrelativistic limit we have
 \begin{subequations}
 \begin{gather}
 \tfrac{1}{m}\left<P^\mu\right>_{\hat{\rho}}\to\delta^{\mu}_{0},\\
 \left<W_0\right>_{\hat{\rho}}\to 0,\\
 \left<S_{0\mu}\right>_{\hat{\rho}}\to 0,\\
 \left<S_{ij}\right>_{\hat{\rho}}\to-\varepsilon_{ijk}
 \tfrac{1}{m}\left<W^k\right>_{\hat{\rho}}.
 \end{gather} 
 \label{nonrelativistic}
 \end{subequations}

The formalism we have introduced above can be straightforward
generalized to the multiparticle case. As an example we shall discuss
briefly the reduced spin density matrix for two massive particles.
Two--particle Hilbert space is spanned by vectors
$\ket{\alpha,k}\otimes\ket{\beta,p}$, where $\ket{\alpha,k}$ is
defined by Eq.\ (\ref{states_new}). Therefore we define the two-particle
unnormalized reduced density matrix as follows:
 \begin{equation}
 \theta_{\alpha^\prime\beta^\prime,\alpha\beta}= \int d\mu(k)\,
 d\mu(p)\, \bra{\alpha,k}\otimes\bra{\beta,p}\hat{\rho}
 \ket{\alpha^\prime,k}\otimes\ket{\beta^\prime,p},
 \label{reduced_two_particle} 
 \end{equation}
where $\hat{\rho}$ denotes the complete two--particle density matrix. It
is obvious that the matrix (\ref{reduced_two_particle}) is Hermitian,
positive--semidefinite and can be easily normalized similarly like in
the one--particle case. Moreover, in the case of two spin $1/2$
particles we define
 \begin{equation}
 \Omega = \theta (\gamma^0\otimes\gamma^0).
 \label{Omega_two_particles}
 \end{equation}

\section{Particle with a sharp momentum}

Now let us discuss the case of the particle with a sharp momentum, say 
$\vec{q}$, and polarization determined by the Bloch vector
$\boldsymbol{\xi}$, $|\boldsymbol{\xi}|\le1$, i.e.,  we assume that
the complete density matrix has the following matrix elements
 \begin{multline}
 \bra{k,m,s,\tau}\hat{\rho}\ket{p,m,s,\lambda} \\ 
 =\frac{2q^0}{\delta^3(\vec{0})} \delta^3(\vec{k}-\vec{q})
 \delta^3(\vec{p}-\vec{q})
 \frac{1}{2}(I-\boldsymbol{\xi}\cdot\boldsymbol{\sigma})_{\tau\lambda}.
 \label{matrix_example}
 \end{multline}
Of course the normalization factor $\tfrac{1}{\delta^3(\vec{0})}$ should be
understood as the result of the proper regularization procedure. 
Now, using Eqs.\ (\ref{reduced_density_new}) and
(\ref{additional_relations}) we can find the corresponding 
matrix $\Omega$. We have
 \begin{equation}
 \Omega = \frac{1}{4} \left( \frac{q\gamma}{m}+I \right) 
 \left(I+2\gamma^5\frac{w\gamma}{m} \right),
 \label{Omega_sharp}
 \end{equation} 
where the four-vector $w^\mu=\left<W^\mu\right>_{\hat{\rho}}$ is given
in this case by
 \begin{equation}
 w^0=\frac{\vec{q}\cdot\boldsymbol{\xi}}{2},\quad
 \vec{w}=\frac{1}{2}\left(m\boldsymbol{\xi} + 
 \frac{\vec{q}(\vec{q}\cdot\boldsymbol{\xi})}{q^0+m}\right),
 \label{Pauli_Lubanski_sharp}
 \end{equation}
i.e., $w$ is obtained from $(0,m\frac{\boldsymbol{\xi}}{2})$ by
applying the Lorentz boost $L_q$. It should also be noted that $w^\mu
q_\mu=0$.  
The matrix (\ref{Omega_sharp}) is known in the literature as the spin
density matrix for Dirac particle
\cite{cab_BLP1968}.

Now, to connect the density
matrix introduced above with some macroscopic experiments like the
Stern--Gerlach one  
let us consider a charged particle with sharp
momentum moving in the external electromagnetic field. We
assume that the giromagnetic ratio $g=2$. 
The momentum and polarization of such a particle vary in time, thus
they can be regarded as functions of its proper time $\tau$:
 \begin{equation}
 q=q(\tau),\quad \boldsymbol{\xi}=\boldsymbol{\xi}(\tau),
 \end{equation}
The expectation value of the operators representing the
spin and the momentum will necessarily follow the same time dependence
as one would obtain from the classical equations of motion
\cite{cab_BMT1959,cab_Corben1961,cab_Anderson1967,cab_CM1994}: 
  \begin{align}
 \frac{dq^\mu}{d\tau} = & \frac{e}{m} F^{\mu}_{\phantom{\mu}\beta}
 q^\beta + \frac{\zeta}{m^2}q^\beta w^\nu
 \partial_\nu\tilde{F}^{\mu}_{\phantom{\mu}\beta} \nonumber\\
 & \mbox{}+\frac{\zeta^2}{m}
 \tilde{F}^{\mu}_{\phantom{\mu}\beta}\big[ 
 F^{\beta}_{\phantom{\beta}\nu}w^\nu + \frac{1}{m^2}q^\beta 
 (w^\sigma F_{\sigma\alpha} q^\alpha )
 \big], \label{trajectory}\\
 \frac{dw^\mu}{d\tau} = & \zeta \big[ F^{\mu}_{\phantom{\mu}\nu}
 w^\nu + \frac{1}{m^2}q^\mu 
 (w^\sigma F_{\sigma\alpha} q^\alpha )\big],  \label{eq_motion_w}
 \end{align}
where $e$ denotes the charge of the particle, $m$ its mass, $\zeta$ is
the proportionality constant between the magnetic moment of the
particle $\mu^\alpha$ and $w^\alpha$ i.e.\
$\mu^\alpha=\frac{\zeta}{m}w^\alpha$\footnote{For the particle with
  charge $e$ and giromagnetic ratio $g$
  $\zeta=\frac{ge}{2m}$\cite{cab_CM1994}.} and  
$F_{\mu\nu}$ is the tensor of the external electromagnetic field,
$\tilde{F}_{\alpha\beta} =
\frac{1}{2}\varepsilon_{\alpha\beta}^{\phantom{\alpha\beta}\mu\nu}F_{\mu\nu}$.
Eq.\ (\ref{eq_motion_w}) describes
Thomas precession of the spin vector in the electromagnetic field
\cite{cab_BMT1959} while Eq.\ (\ref{trajectory}) allows one to determine
the trajectory of the spinning particle moving in the 
electromagnetic field $F_{\mu\nu}$. The slow motion limit of the above
equations takes the well--known form \cite{cab_CM1994}
 \begin{gather}
 \frac{d\vec{q}}{dt} = \frac{e}{m} \vec{q}\times\vec{B}
 +\frac{\zeta}{2}\boldsymbol{\xi}\cdot \nabla\vec{B}, \label{Stern_Gerlach_1}\\
 \frac{d\boldsymbol{\xi}}{dt} = \zeta \boldsymbol{\xi} \times
 \vec{B}, \label{Stern_Gerlach_2}
 \end{gather}
where we assumed that the electric
component of the electromagnetic field is equal to zero.
Eqs.\ (\ref{Stern_Gerlach_1}--\ref{Stern_Gerlach_2}) describe forces
acting on the particle in the Stern--Gerlach experiment, therefore we
can really identify $\boldsymbol{\xi}$ with the polarization of the
particle. 

In this simple case of the monochromatic particle we can also
calculate explicitly the von Nuemann entropy of the reduced density
matrix. The matrix $\Omega$ in the rest frame of the particle
can be written as
 \begin{equation}
 \Omega_0 = \tfrac{1}{2} \begin{pmatrix}
 1 & 1 \\ 1 & 1
 \end{pmatrix}\otimes \tfrac{1}{2}(I+
 \boldsymbol{\xi}\cdot\boldsymbol{\sigma}).  
 \label{matrix_rest}
 \end{equation} 
To calculate entropy we have to use the normalized density matrix
$\tilde{\theta}_0$, but in this particular case
$\tilde{\theta}_0=\Omega_0$. Thus the von Neumann entropy of the state
(\ref{matrix_rest}) is equal to
 \begin{equation}
 S_{\tilde{\theta}_0}=-\frac{1}{2}\left((1+|\boldsymbol{\xi}|)
 \ln{\frac{1+|\boldsymbol{\xi}|}{2}}+ (1-|\boldsymbol{\xi}|)
 \ln{\frac{1-|\boldsymbol{\xi}|}{2}}
 \right). 
 \label{entropy}
 \end{equation}
Now, to find the entropy in the arbitrary Lorentz frame we apply to
the matrix $\tilde{\theta}_0$ the Lorentz transformation
(\ref{transf_normalized}) with $\rep{D}(\Lambda)$ given by
(\ref{representation}) and
we find that entropy of the corresponding
reduced density matrix $\tilde{\theta}^\prime$ is given by
(\ref{entropy}) too, i.e.,
$S_{\tilde{\theta}_0}=S_{\tilde{\theta}^{\prime}}$. Therefore for
a particle with the sharp momentum the 
entropy of the reduced density matrix does not change under Lorentz
transformations. However, in the case of an arbitrary momentum
distribution, the entropy of the reduced density matrix
$\tilde{\theta}$ is not in general Lorentz--invariant.

\section{Spin operator}

In the next section we will use our formalism to
calculate the Einstein--Podolsky--Rosen--Bohm (EPR--Bohm)
correlation function. Thus we have to introduce the spin operator for a
relativistic massive particle. The choice is not obvious since
in the discussion of relativistic EPR--Bohm experiments various spin 
operators have been used
\cite{cab_ALMH2003,cab_Czachor1997_1,cab_RS2002,%
cab_LCY2004,cab_LD2003,cab_TU2003_1,cab_TU2003_2}.
However our previous considerations (Eqs.\
(\ref{Pauli_Lubanski_sharp})--(\ref{Stern_Gerlach_2})) 
as well as the classical definition of the relativistic spin
\cite{cab_Anderson1967} suggest that the best
candidate for the spin operator is
 \begin{equation}
 \Hat{\vec{S}}= \frac{1}{m}\left(\Hat{\vec{W}} -
   \Hat{W}^0\frac{\Hat{\vec{P}}}{\Hat{P}^0+m} \right), 
 \label{spin}
 \end{equation}
which corresponds to the classical polarization vector
$\boldsymbol{\xi}$ (precisely to $\boldsymbol{\xi}/2$) in
Eq.\ (\ref{Pauli_Lubanski_sharp}). This operator is also used in 
quantum field theory \cite{cab_BLT1969}. It
fulfills the following standard commutation relations:
\begin{subequations}
 \begin{gather}
 [\Hat{J}^i,\Hat{S}^j]=i\varepsilon_{ijk}\Hat{S}^k,\\
 [\Hat{S}^i,\Hat{S}^j]=i\varepsilon_{ijk}\Hat{S}^k,  
 \label{spin_commutator} \\ 
 [\Hat{P}^\mu,\Hat{S}^j]=0,
 \end{gather}%
\label{spin_all_commutators}%
\end{subequations}%
which should be satisfied for the spin operator.
Here $\Hat{J}^i=\frac{1}{2}\varepsilon_{ijk}\Hat{J}^{jk}$ and one can
show that it is the only operator which is a linear function of
$\Hat{W}^\mu$ and fulfills relations (\ref{spin_all_commutators})
\cite{cab_BLT1969}. 

Therefore the operator corresponding to the spin projection along
arbitrary direction $\vec{n}$ ($\vec{n}^2=1$)
in the representation of gamma matrices (\ref{gamma_explicit})
reads explicitly
 \begin{multline}
 \vec{n}\cdot\Hat{\vec{S}} = \frac{1}{2m}\left\{
  \Hat{P}^0 
 \begin{pmatrix} \vec{n}\cdot\boldsymbol{\sigma} & 0 \\
 0 & \vec{n}\cdot\boldsymbol{\sigma} \end{pmatrix}\right. \\
  - i  
 \begin{pmatrix} (\vec{n}\times\Hat{\vec{P}})\cdot\boldsymbol{\sigma} & 0 \\
 0 & -(\vec{n}\times\Hat{\vec{P}})\cdot\boldsymbol{\sigma}
 \end{pmatrix}\\
 \left.
  -\frac{\vec{n}\cdot\Hat{\vec{P}}}{\Hat{P}^0+m}
 \begin{pmatrix} \Hat{\vec{P}}\cdot\boldsymbol{\sigma} & 0 \\
 0 & \Hat{\vec{P}}\cdot\boldsymbol{\sigma} \end{pmatrix} \right\},
 \label{nS}
 \end{multline}
where we have used Eqs.\ (\ref{Pauli_Lubanski_explicit}).

Eq.\ (\ref{spin_commutator}) implies that eigenvalues of the operator
$\vec{n}\cdot\Hat{\vec{S}}$ are integers or half--integers.  
As one can easily check by direct calculation the eigenvalues of the
operator (\ref{nS}) are equal to $\pm\frac{1}{2}$. This observation
supports our choice of the operator $\Hat{\vec{S}}$ as the spin
operator. 

Now we want to express the average of the spin operator (\ref{spin})
in terms of the reduced matrix $\Omega$. One can check that in an
arbitrary state $\Hat{\rho}$
 \begin{equation}
 \left<(\Hat{P}^0+m)\Hat{\vec{S}}\right>_{\Hat{\rho}} = \frac{m}{2}
 \text{Tr}\,  
 \left(\Omega \boldsymbol{\gamma} \gamma^5 (I+\gamma^0) \right). 
 \label{average_spin_times_P0}
 \end{equation}
Thus a reasonable choice for the normalized average of the spin
component is  
 \begin{equation}
 \boldsymbol{\Sigma}  = 
 \frac{\left<(\Hat{P}^0+m)\Hat{\vec{S}}\right>_{\Hat{\rho}}}%
 {\left<(\Hat{P}^0+m)\right>_{\Hat{\rho}}} 
  = \frac{\text{Tr}\, 
 \left(\Omega \boldsymbol{\gamma} \gamma^5 (I+\gamma^0) \right)}%
 {2\text{Tr}\,\left(\Omega(I+\gamma^0)\right)}.
 \label{spin_average_sharp}
 \end{equation}
When $\Hat{\rho}(k)$ describes a particle with a sharp momentum $k$ the
normalized average is simply the average of $\Hat{S}$, i.e.,
inserting reduced density matrix $\Omega$ (\ref{Omega_sharp}) into  
(\ref{spin_average_sharp}) we get
 \begin{equation}
 \boldsymbol{\Sigma}  = \left< \Hat{\vec{S}} \right>_{\Hat{\rho}(k)}
 =  \frac{\boldsymbol{\xi}}{2}.
 \label{spin_average_sharp_sigma}
 \end{equation}

It should also be noted that in the nonrelativistic limit we recover
the result 
(\ref{spin_average_sharp_sigma}) for an arbitrary state $\Hat{\rho}$  
 \begin{equation}
 \boldsymbol{\Sigma} = \left< \Hat{\vec{S}}\right>_\rho =
 \frac{\boldsymbol{\xi}}{2}. 
 \end{equation}

\section{Quantum correlations}

Using the formalism introduced above, we now calculate the
correlation between
measurements of spin components performed by two observers, \textsf{A}
and \textsf{B}, along two
arbitrary directions, $\vec{a}$ and $\vec{b}$, respectively.  We 
consider the simplest situation in which 
both observers are at rest with respect to a certain inertial frame of
reference ${\mathcal{O}}$. We assume also that both measurements
are performed simultaneously in the frame ${\mathcal{O}}$.

We calculate the EPR--Bohm correlation function in the pure state
of two particles with sharp momenta
 \begin{equation}
 \ket{\psi} = \sum_{\alpha\beta} c_{\alpha\beta}
 \ket{\alpha,k}\otimes\ket{\beta,p}.
 \label{state_psi}
 \end{equation}
The corresponding reduced density matrix (\ref{Omega_two_particles})
has the following form:
 \begin{multline}
 \Omega_{\alpha\beta,\alpha^\prime\beta^\prime}^{\psi} = 
 \frac{4 k^0 p^0 (\delta^3(\vec{0}))^2}{\left<\psi|\psi\right>} \\
 \left[v(k)\bar{v}(k) \gamma^0 C^* \gamma^0
   \big(v(p)\bar{v}(p)\big)^T \right]_{\alpha\beta} \\
 \left[ \big(v(k)\bar{v}(k) \big)^T C v(p)\bar{v}(p)
 \right]_{\alpha^\prime\beta^\prime},
 \end{multline}
where the matrix $C=(c_{\alpha\beta})$ determines the state
(\ref{state_psi}) while $v(k)\bar{v}(k)$ and $v(p)\bar{v}(p)$ are given by 
(\ref{normalization_b}).

Observers \textsf{A} and \textsf{B} use observables
$2\vec{a}\cdot\Hat{\vec{S}}\otimes I$ and
$I\otimes2\vec{b}\cdot\Hat{\vec{S}}$, 
respectively ($\vec{a}^2=\vec{b}^2=1$). Thus the correlation function has
the form (see Eqs.\ (\ref{average_spin_times_P0})
and (\ref{two_particles_tensor_observable}))
 \begin{align}
 \mathcal{C}(\vec{a},\vec{b})&  =4 
 \frac{\left<(\Hat{P}^0+m) \vec{a}\cdot\Hat{\vec{S}} \otimes 
 (\Hat{P}^0+m) \vec{b}\cdot\Hat{\vec{S}} \right>_\psi}{\left<
 (\Hat{P}^0+m)\otimes(\Hat{P}^0+m)\right>_\psi} \nonumber\\
 & = \frac{\text{Tr}\,\left[
     \Omega^\psi\Big((\vec{a}\cdot\boldsymbol{\gamma}\gamma^5(I+\gamma^0)) 
 \otimes(\vec{b}\cdot\boldsymbol{\gamma}\gamma^5(I+\gamma^0))\Big)\right]}%
 {\text{Tr}\,\left[ \Omega^\psi\Big((\gamma^0+I)\otimes
 (\gamma^0+I)\Big)\right]}\nonumber \\
 & =4
 \frac{\bra{\psi}\vec{a}\cdot\Hat{\vec{S}}\otimes \vec{b}\cdot\Hat{\vec{S}}
 \ket{\psi}}{\left<\psi|\psi\right>}.
 \end{align}
After some algebra we find that
 \begin{multline}
 \mathcal{C}(\vec{a},\vec{b}) = \\
 \frac{\text{Tr}\left\{ \left(\vec{b}\cdot\vec{S}(p)v(p)\bar{v}(p)\gamma^0\right) C^\dag
   \left(\vec{a}\cdot\vec{S}(k)v(k)\bar{v}(k)\gamma^0\right)^T C\right\}}{%
 \text{Tr}\left\{\left(v(p)\bar{v}(p)\gamma^0\right) C^\dag
   \left(v(k)\bar{v}(k)\gamma^0\right)^T C\right\}}, 
 \end{multline}
where matrices $\vec{a}\cdot\vec{S}(k)$ and
$\vec{b}\cdot\vec{S}(p)$ have the same form as (\ref{nS}) with $\vec{n}$,
$\Hat{P}$ equal to $\vec{a}$, $k$ and $\vec{b}$, $p$ respectively. 
Now,
for the sake of simplicity, we  
specify the state $\ket{\psi}$. We choose 
 \begin{equation}
 C=a \begin{pmatrix} \sigma_2 & 0 \\ 0 & \sigma_2 \end{pmatrix},  
 \label{state_psi_singlet}
 \end{equation}
where $a$ is a normalization constant. This choice is rather
natural because the state described by Eqs.\ (\ref{state_psi}),
(\ref{state_psi_singlet}) has the same form for all inertial
observers, namely
 \begin{equation}
 U(\Lambda)\otimes U(\Lambda) \ket{\psi} = \sum_{\alpha\beta}
 c_{\alpha\beta} \ket{\alpha,\Lambda k}\otimes \ket{\beta,\Lambda p},
 \end{equation}
where we used Eq.\ (\ref{representation}). Moreover in the center of mass
frame it is an ordinary singlet state. 

Now, provided that $C$ is given by Eq.\ (\ref{state_psi_singlet}), after
straightforward calculation we arrive at
 \begin{multline}
 \mathcal{C}(\vec{a},\vec{b}) = -\vec{a}\cdot\vec{b} 
 + \frac{(\vec{k}\times\vec{p})}{m^2+kp}\cdot
 \bigg((\vec{a}\times\vec{b}) \\
   +\frac{(\vec{a}\cdot\vec{k})(\vec{b}\times\vec{p}) -
     (\vec{b}\cdot\vec{p})(\vec{a}\times\vec{k})}{(k^0+m)(p^0+m)}\bigg).
 \end{multline}
We see that the correction to the nonrelativistic correlation function
$\Delta{\mathcal{C}} = \mathcal{C}(\vec{a},\vec{b}) -
{\mathcal{C}}_{\text{nonrel}}=\mathcal{C}(\vec{a},\vec{b}) +
\vec{a}\vec{b}$ is of order $\beta^2$, where $\beta=\frac{v}{c}$,
$v$ denotes the velocity of the particle, and $c$ the velocity of light.
Let us note first that when momenta of both particles are parallel or
antiparallel the 
correlation function has the same form as in the nonrelativistic
case. This result differs from Czachor's results\footnote{The
  Czachor's result can be obtained in our framework by calculating the
  appriopriately normalized average of
  \mbox{$\vec{a}\cdot\Hat{\vec{W}}\otimes\vec{b}\cdot\Hat{\vec{W}}$}.} 
\cite{cab_Czachor1997_1}. The reason is that we use a different, and in our
opinion more adequate, spin operator.

\begin{figure}
\includegraphics[width=\columnwidth]{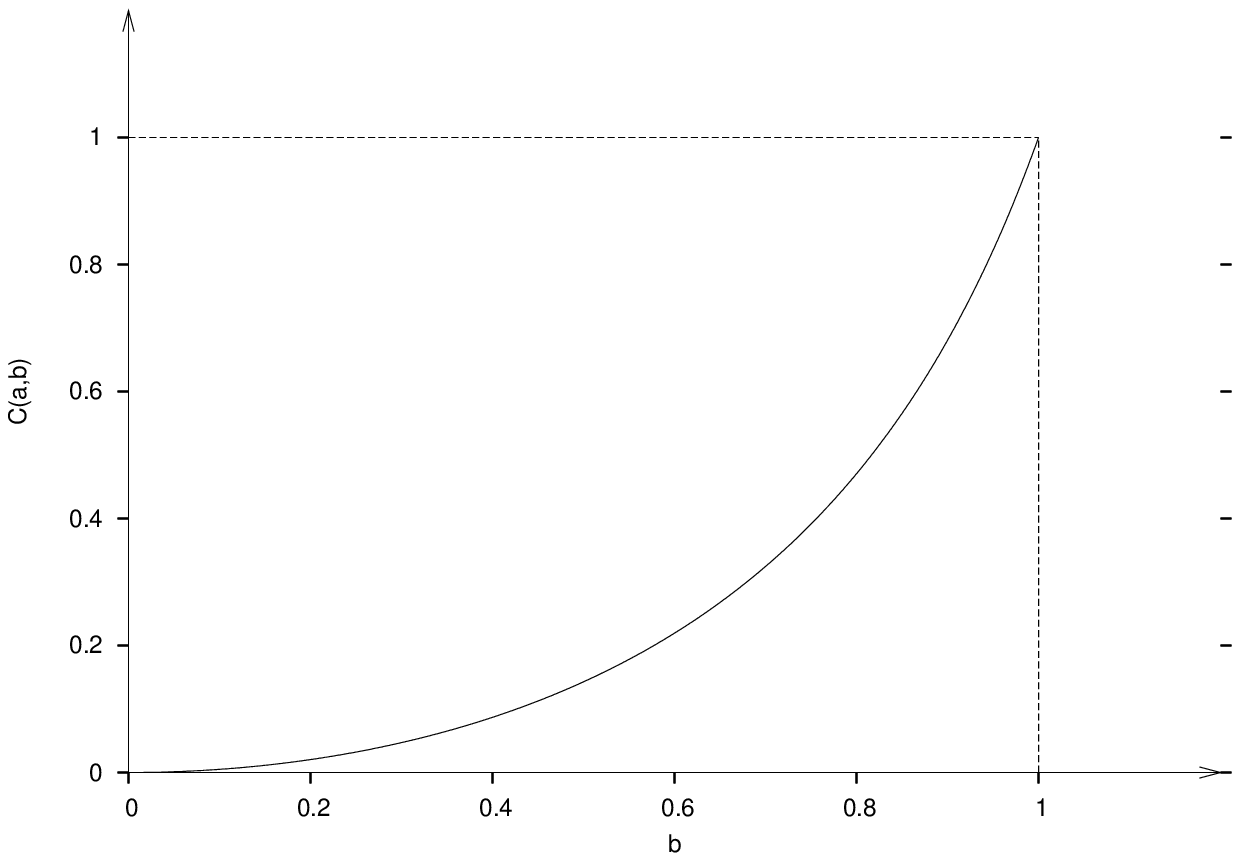}
\caption{Correlation function in the case when 
$\vec{a}\perp\vec{b}$, $|\vec{k}|=|\vec{p}|$ and
$\vec{a}\parallel\vec{k}$, $\vec{b}\parallel\vec{p}$ or
$\vec{a}\perp\vec{k}$, $\vec{b}\perp\vec{p}$ (Eq.\
  (\ref{correlation_only_relativistic}))}
\label{rys}
\end{figure}

Now let us consider the configuration in which the nonrelativistic
correlation vanishes ($\vec{a}\perp\vec{b}$). For simplicity let us
assume also that 
$|\vec{k}|=|\vec{p}|$ and $\vec{a}\parallel\vec{k}$,
$\vec{b}\parallel\vec{p}$ or $\vec{a}\perp\vec{k}$,
$\vec{b}\perp\vec{p}$. In such fixed configurations the correlation
function has the very simple form 
 \begin{equation}
 \mathcal{C}(\vec{a},\vec{b}) = \Delta{\mathcal{C}} =
 \frac{p_{0}^{2}-m^2}{p_{0}^{2}+m^2}= 
 \frac{\beta^2}{2-\beta^2}.
 \label{correlation_only_relativistic}
 \end{equation}
Dependence of the above correlation on $\beta$ is depicted in
Fig.\ \ref{rys}. Notice that (\ref{correlation_only_relativistic}) was
also obtained by Czachor \cite{cab_Czachor1997_1} but for a different
configuration.

\section{Conclusions}

To conclude, we have constructed a Lorentz--covariant reduced spin
density matrix for a single massive particle. 
It contains not only information about \textit{average} polarization
of the particle but also information about its \textit{average}
kinematical state. We have also showed that this matrix has the proper
nonrelativistic limit.

Our results shows
that we  can define a Lorentz--covariant finite--dimensional matrix
describing polarization of a massive particle.
However in the relativistic case (contrary to the
nonrelativistic one) we cannot completely separate kinematical 
degrees of freedom if we want to construct a finite-dimensional
covariant description of the polarization degrees of freedom.

With help of our covariant formalism we have also calculated
the correlation function in the EPR--Bohm type
experiment with massive relativistic particles. 
We have showed that relativistic
correction $\Delta{\mathcal{C}}$
to the nonrelativistic correlation function
${\mathcal{C}}_{\text{nonrel}}=-\vec{a}\cdot\vec{b}$ vanishes when
momenta of both particles are 
parallel or antiparallel, i.e., in the standard configuration of
EPR--Bohm type experiments. We have found also the configurations in
which the nonrelativistic correlation vanishes while the relativistic
correction $\Delta{\mathcal{C}}$ survives and is  of order
$\beta^2$ (Eq.\ (\ref{correlation_only_relativistic})). 

\begin{acknowledgments}
The authors thank Marek Czachor for interesting discussions.
This paper has been partially supported by the Polish Ministry of
Scientific Research and Information Technology under Grant No.\ 
PBZ-MIN-008/P03/2003 and partially by the University of Lodz grant.
\end{acknowledgments}

\appendix
\section{Dirac matrices}
In this paper we use the following conventions.
Dirac matrices fulfills the condition
$\gamma^\mu\gamma^\nu+\gamma^\nu\gamma^\mu=2g^{\mu\nu}$ where
$g^{\mu\nu}=\text{diag}(1,-1,-1,-1)$ denotes Minkowski metric tensor;
moreover we adopt the convention $\varepsilon^{0123}=1$.  We use the
following explicit representation of gamma matrices:
 \protect\begin{equation}
 \gamma^0=\left(\protect\begin{array}{cc}
 0 & I \\ I & 0
 \protect\end{array}\right),\quad 
 {\boldsymbol{\gamma}}=\left(\protect\begin{array}{cc}
 0 & -{\boldsymbol{\sigma}} \\ {\boldsymbol{\sigma}} & 0
 \protect\end{array}\right),\quad \gamma^5= 
 \left(\protect\begin{array}{cc} 
 I & 0 \\ 0 & -I
 \protect\end{array}\right),
 \label{gamma_explicit}
 \protect\end{equation} 
where $\boldsymbol{\sigma}=(\sigma_1,\sigma_2,\sigma_3)$ and
$\sigma_i$ are standard Pauli matrices. 

\section{Useful formulas}

The matrix (\ref{amplitude}) is normalized as follows
\protect\begin{subequations}
 \protect\begin{gather}
 \bar{v}(k) v(k) = I, \\
 v(k) \bar{v}(k) = \tfrac{1}{2m}(k\gamma+mI),\label{normalization_b}
 \protect\end{gather}
\label{normalization}\unskip
\protect\end{subequations}\unskip
where $\bar{v}(k)=v^\dag(k)\gamma^0$. 
Moreover it can be verified that
it fulfills the following relation 
 \protect\begin{equation}
 \bar{v}(k)\gamma^\mu v(k) = \frac{k^\mu}{m}I.
 \label{additional_relations} 
 \protect\end{equation}
Vectors $\ket{\alpha,k}$ fulfill the orthogonality relation:
 \begin{equation}
 \left<\alpha,k|\beta,p\right> = 2 k^0 \delta^3(\vec{k}-\vec{p})
 \left(v(k)v^\dag(k)\right)_{\beta\alpha},
 \label{orthogonality_new}
 \end{equation}
and one can check that
 \begin{gather}
 I=\sum_{\alpha\beta}\int d\mu(k) \gamma^{0}_{\beta\alpha}
 \ket{\alpha,k}\bra{\beta,k}, \\
 \Big(v(k)\bar{v}(k)\Big)_{\alpha\beta}\ket{\beta,k} =
 \ket{\alpha,k}. 
 \end{gather}
In the representation of gamma matrices (\ref{gamma_explicit})
we have
\begin{subequations}
 \begin{gather}
 \Hat{W}^0 = \frac{1}{2} 
 \begin{pmatrix} \Hat{\vec{P}}\cdot\boldsymbol{\sigma} &  0 \\
 0 & \Hat{\vec{P}}\cdot\boldsymbol{\sigma} \end{pmatrix}, \\
 \Hat{\vec{W}} = \frac{1}{2} \Hat{P}^0 
 \begin{pmatrix} \boldsymbol{\sigma} & 0 \\
 0 & \boldsymbol{\sigma} \end{pmatrix}
 - \frac{i}{2} 
 \begin{pmatrix} \Hat{\vec{P}}\times\boldsymbol{\sigma} & 0 \\ 
 0 & - \Hat{\vec{P}}\times\boldsymbol{\sigma} \end{pmatrix}.
 \end{gather}%
\label{Pauli_Lubanski_explicit}%
\end{subequations}
It can be also checked, that when
\begin{subequations}
 \begin{gather}
 \left<\Hat{F}\right>_\rho = \text{Tr}\,(\Omega f),\\
 \left<\Hat{G}\right>_\rho = \text{Tr}\,(\Omega g),
 \end{gather}%
\label{tensor_observables}%
\end{subequations}%
we have
 \begin{equation}
 \left<\Hat{F}\otimes\Hat{G}\right>_\rho = \text{Tr}\,(\Omega
 (f\otimes g)),
 \label{two_particles_tensor_observable}
 \end{equation}
where in Eqs.\ (\ref{tensor_observables}) and
(\ref{two_particles_tensor_observable}) $\rho$ and $\Omega$ are
complete and reduced density matrices for one and two particles,
respectively.



\end{document}